# Measuring What a Specification Determines: A Formal Semantic-Block Model and an Execution-Judged Benchmark

Oleg Grynets
EPAM Systems
McLean, Virginia, USA
oleg_grynets@epam.com

Dmytro Kostetskyi
EPAM Systems
Lviv, Ukraine
dmytro_kostetskyi@epam.com

Vasyl Lyashkevych
EPAM Systems
Lviv, Ukraine
vasyl_lyashkevych@epam.com

***Abstract*—Specification-driven development (SDD) increasingly treats the specification as the primary operational artifact consumed by large language model (LLM) development agents. This work introduces a formal semantic-block model for specifications and an execution-judged benchmark for evaluating specification quality independently of model capability. A specification is represented as a structure comprising semantic blocks, dependency relations, block-owned rules, decision points, and explicitly open questions, subject to four machine-checkable well-formedness conditions: acyclicity, single ownership, constraint domination, and totality or ambiguity-stop. Determinacy is defined model-theoretically as agreement among all conforming implementations and is estimated empirically through convergence across independent implementers. The model is instantiated on an Oracle-to-PostgreSQL migration specification containing 18 blocks and 19 dependency edges. Computational validation shows that the five-layer decomposition reduces mean per-task context by approximately 71% through dependency closures, covers 85.5% of the study-defined Oracle construct taxonomy with all identified gaps triaged, is not Pareto-dominated by the tested alternative partitions, and is recovered at the 99.9th percentile from citation-derived edges not used to define the original structure. The benchmark keeps the implementer panel fixed, includes a mandatory no-specification control arm, and uses PostgreSQL 16 and a live Oracle instance as deterministic execution judges. On 75 execution-verified units, the full specification increases clean-loading output from 72.0% to 97.3%, whereas aggregate cross-implementer agreement changes only from 83.1% to 83.8%. No specification-pinned decision on the evaluated 75-unit decision surface shows a measurable agreement gain over the no-specification baseline, and structured row-level comparison against live Oracle yields 19 correct results out of 42 probes in both conditions. Six designed studies, including three pre-registered manipulations and three diagnostic analyses, further examine specification effects. Repeated runs on a 25-unit subsample reveal an empirical variability floor with a median arm-delta spread of 14.4 percentage points. The results support determinacy as a formal concept but not as a standalone empirical quality metric for the evaluated contemporary LLM implementers. In the studied migration setting, the specification improves executability and checkability and supports gap localization, while aggregate agreement and structured-data correctness do not improve measurably.**

## I. Introduction

LLMs are increasingly used throughout software engineering [1], [2]. In SDD, business requirements, technical constraints, architecture decisions, and acceptance criteria become operational inputs to LLM-based development agents. Consequently, the specification becomes a knowledge boundary between engineering intent and generated software rather than merely supporting documentation. Specification-based reengineering, architecture-aware representations, and security-oriented SDD further demonstrate the growing role of specifications as intermediate knowledge artifacts [3]–[6].

This shift creates a measurement problem. Conventional specification qualities such as completeness, consistency, traceability, and unambiguity do not establish how independent implementers will actually behave. A rule may be present but ignored; implementers may agree because they share an industry default rather than because the specification determined the decision; compliant output may still be incorrect; and executable code may fail to preserve source-system behavior. Thus, "Specification presence ≠ Rule consumption ≠ Compliance ≠ Behavioral correctness".

Related work reinforces this distinction. Security-oriented studies show that specification content can affect observable security behavior [6]–[8], while state- and evolution-oriented models show that software quality is multidimensional and context-dependent [9]–[11]. Execution-based benchmarks likewise demonstrate that generated software should be judged by observable behavior rather than textual similarity alone [12]–[14]. However, existing benchmarks primarily score models under fixed tasks. They do not isolate the specification as the experimental treatment or distinguish specification-induced agreement from decisions implementers would make without it.

This paper addresses that gap through three research questions:

RQ1 — Structure. What minimal formal structure allows a multi-team, multi-agent specification to be authored, machine-checked, and consumed in task-specific slices?

RQ2 — Quality. Can specification quality be measured operationally through independent implementations rather than asserted through document review?

RQ3 — Validity. Does cross-implementer convergence remain a useful estimator of specification determinacy when evaluated at scale against a no-specification control?

We define a specification as:

$$S = (B, \prec, \rho, \Delta, Q), \quad (1)$$

where $B$ denotes semantic blocks, $\prec$ their dependencies, ρ block-owned rules, Δ decision points, and $Q$ explicitly unresolved questions. Four machine-checkable conditions enforce acyclicity, single ownership, constraint domination, and totality or ambiguity-stop. The instantiated "Oracle 19c

→ PostgreSQL 16" specification contains 18 blocks and 19 dependency edges. Its dependency closures reduce mean task context by approximately 71%.

Next, we develop a benchmark based on execution assessment, where the specification—rather than the model itself—serves as the key factor. The executors and test units remain constant across all test variants. A control group without a specification is mandatory. The context is automatically generated from the specification graph. The combination of PostgreSQL 16 and a live Oracle instance ensures deterministic execution and behavioral assessment. The evaluation process encompasses an execution-filtered corpus of 981 Oracle–PostgreSQL pairs, a fixed sample of 75 units, and a separate corpus of 20 units for behavioral analysis.

The results substantially qualify convergence-based specification quality. A full specification increases clean-loading output from 72.0% to 97.3%, yet aggregate three-implementer agreement changes only from 83.1% to 83.8%, and structured-data correctness remains 19 of 42 probes in both control and specification arms. Rule placement matters more: identical wording yields compliance of 23% as a trailing paragraph and 42% when placed in the dominant decision matrix, while duplicating the rule lowers it to 34%. We also observe a rule whose compliance makes the migration worse, and a specification-consumption protocol that produces no measurable improvement. Finally, repeated identical runs reveal a 14.4 percentage-point median spread in the arm delta, forcing the retirement of several smaller earlier claims. The contributions are:

- a formal semantic-block specification model with four machine-checkable well-formedness conditions;
- computational validation of its decomposition and dependency-based context reduction;
- an execution-judged benchmark that isolates the specification through a fixed implementer panel and mandatory no-specification control;
- empirical separation of agreement, compliance, executability, and behavioral correctness;
- evidence that presentation can affect rule uptake more strongly than repetition or meta-level consumption instructions;
- an empirical characterization of run-to-run variability showing that convergence alone is insufficient for judging specification quality.

Overall, the results support determinacy as a formal notion but not as a standalone empirical quality metric for the evaluated contemporary LLM implementers. In the evaluated setting, the specification demonstrably improves executability, checkability, and gap localization, but not aggregate agreement or structured-data correctness.

## II. Related Work

LLMs are increasingly applied across requirements engineering, code generation, testing, maintenance, and software evolution [1], [2]. SDD approaches shift attention from model capability toward the engineering artifact that controls generation. In specification-based Code–Text–Code reengineering, a neutral specification acts as an intermediate representation preserving behavioral, structural, dependency, and domain knowledge during software transformation [3]. Architecture-oriented work similarly introduces structured representations connecting requirements, architecture, documentation, and generated code [4]. Practitioner-oriented SDD frameworks and specification-code fidelity studies further emphasize specifications as executable contracts and the need to measure alignment between specifications and generated artifacts [15], [16].

Token-optimization experiments for Oracle→PostgreSQL migration further show that context reduction must preserve semantic information: aggressive compression can improve token efficiency while substantially reducing semantic match [5]. Oracle-to-PostgreSQL migration has additionally been studied through fine-tuned LLM-based transformation pipelines [17]. These findings motivate our dependency-closure approach, which reduces context structurally rather than by rewriting retained specification content.

Our use of multiple implementers is related to N-version programming. Avizienis proposed independently developed versions for fault tolerance [18], while Knight and Leveson showed that independent implementations can nevertheless exhibit correlated failures [19]. This limitation is particularly relevant to LLMs, which may share training distributions and industry conventions. Therefore, high cross-implementer agreement cannot by itself demonstrate specification determinacy. We address this by keeping the implementer panel fixed and introducing a mandatory no-specification control that distinguishes specification-induced agreement from shared defaults.

Specification quality is traditionally associated with completeness, consistency, traceability, and unambiguity [20], while formal methods such as Alloy and TLA+ provide precise representations of structural and behavioral constraints [21], [22]. Our objective differs from full formalization: we formalize the organization and determination properties of specifications that remain consumable as structured technical language. In particular, determinacy concerns whether conforming implementations resolve a decision consistently, while the ambiguity-stop condition requires unresolved decisions to be explicit rather than silently omitted.

Requirements-quality research has operationalized ambiguity, incompleteness, and requirements smells using linguistic and visualization-based techniques [23]–[27]. These methods assess defects in specification text, whereas our benchmark evaluates whether specification decisions alter downstream implementation behavior.

Specification quality is also property-specific. Research on LLM-based software development identifies security risks that can propagate throughout the SDLC [7], while vulnerability has been modeled as a characteristic of software functional state [8]. The companion SDD Benchmark: Security Knowledge Transition operationalizes this problem by evaluating whether security knowledge encoded in a specification survives generation into observable security behavior [6]. These studies support separating specification presence, rule compliance, and downstream correctness rather than treating them as a single quality dimension.

Empirical studies of AI-assisted programming further show that generated or AI-assisted code can introduce security weaknesses even when functional requirements appear satisfied [28], [29].

This multidimensional interpretation is consistent with state- and evolution-oriented software models. Software Functional State models describe software quality as changing throughout the lifecycle [9]; context-aware intelligent monitoring relates system state to contextual decision strategies [10]; and multi-drift monitoring distinguishes configuration, topology, role, policy, architectural, contextual, semantic, goal, and security changes [11]. The present benchmark evaluates a frozen specification rather than specification evolution, but these works establish an important boundary: specification adequacy at one system state does not imply adequacy after system or context changes.

Finally, modern software and database benchmarks increasingly prefer executable evidence over textual similarity. SWE-bench evaluates repository-level changes against executable tasks [12], while execution-based SQL evaluation demonstrates that textual or single-reference matching can misrepresent semantic equivalence [13]. BIRD extends this principle to large-scale database-grounded Text-to-SQL evaluation [14].

Execution- and test-based evaluation has also become central to LLM software engineering, including automated unit-test generation [30], [31], repository-level code completion [32], and increasingly realistic SQL benchmarks such as Spider and Spider 2.0 [33], [34]. Our benchmark adopts the same execution-first principle but changes the experimental object: the implementers are held constant and the specification is the treatment. PostgreSQL execution and differential comparison against live Oracle are then used to separate agreement, compliance, executability, and behavioral correctness.

The resulting gap is specific: prior work studies specification representation [3]–[5], security knowledge transition [6]–[8], software state and evolution [9]–[11], model-centered executable evaluation [12]–[14], and independent implementation [18], [19], but does not jointly measure what behavior is actually caused by a specification. This paper addresses that gap through a formal semantic-block model, a no-specification control, and deterministic execution-based judging.

## III. Semantic-Block Specification Model

The specification under study governs an Oracle 19c → PostgreSQL 16 migration. It contains 18 structured documents and 61,709 tokens, organized into five semantic layers corresponding to successive migration concerns: foundation, data and schema, procedural logic, migration pipeline, and quality/testing. Each block owns a distinct set of decisions and consumes decisions established by its dependencies.

The five layers are:

- 100 — Foundation: project constraints, business goals, security, and architecture decisions;
- 200 — Data and Schema: datatype mapping, tables, indexes, sequences, identity, collation, and charset;
- 300 — Procedural Logic: package decomposition, cursors and loops, exception handling, and transaction control;
- 400 — Migration Pipeline: initial load, CDC replication, cutover, and fallback;
- 500 — Quality and Testing: data integrity, functional testing, and performance benchmarking.

Each document follows a common machine-parsable structure containing status metadata, dependencies, purpose and scope, implementation rules, known pitfalls, open questions, and revision history. Hardening changes carry provenance linking a revised rule to the gap that motivated it. Thus, unresolved decisions and later corrections remain explicitly traceable.

Three principles govern the specification.

**Single ownership**. Each rule belongs to exactly one semantic block; other blocks reference rather than duplicate it. This makes rule ownership and coverage mechanically inspectable.

**Constraint propagation**. Foundation constraints dominate lower-level blocks. In the instantiated model, block 102—System Constraints fixes global restrictions including PostgreSQL version, extension policy, forbidden Oracle constructs, downtime constraints, and encoding.

**Ambiguity-stop**. A decision that is not determined must be represented as an explicit open question. Consumers should not silently replace missing specification knowledge with an unstated assumption. The dependency declarations form a directed graph:

$$G = (B, \prec), \tag{2}$$

where B is the set of semantic blocks and $\prec$ represents dependency. The instantiated graph contains 18 nodes and 19 edges, has density 0.124, a longest path of 7, maximum level width 6, fan-in no greater than 2, and fan-out no greater than 4. Block 101 is the unique root, while block 102 reaches every lower block through the transitive dependency relation.

Fig. 1 illustrates this dependency DAG. The graph is intentionally sparse: every additional edge increases the context that a consumer must retrieve for some downstream task. An edge is therefore included only when one block genuinely consumes a decision owned by another. An edge $b_i \rightarrow b_j$ indicates that block $b_j$ consumes decisions established by block $b_i$.

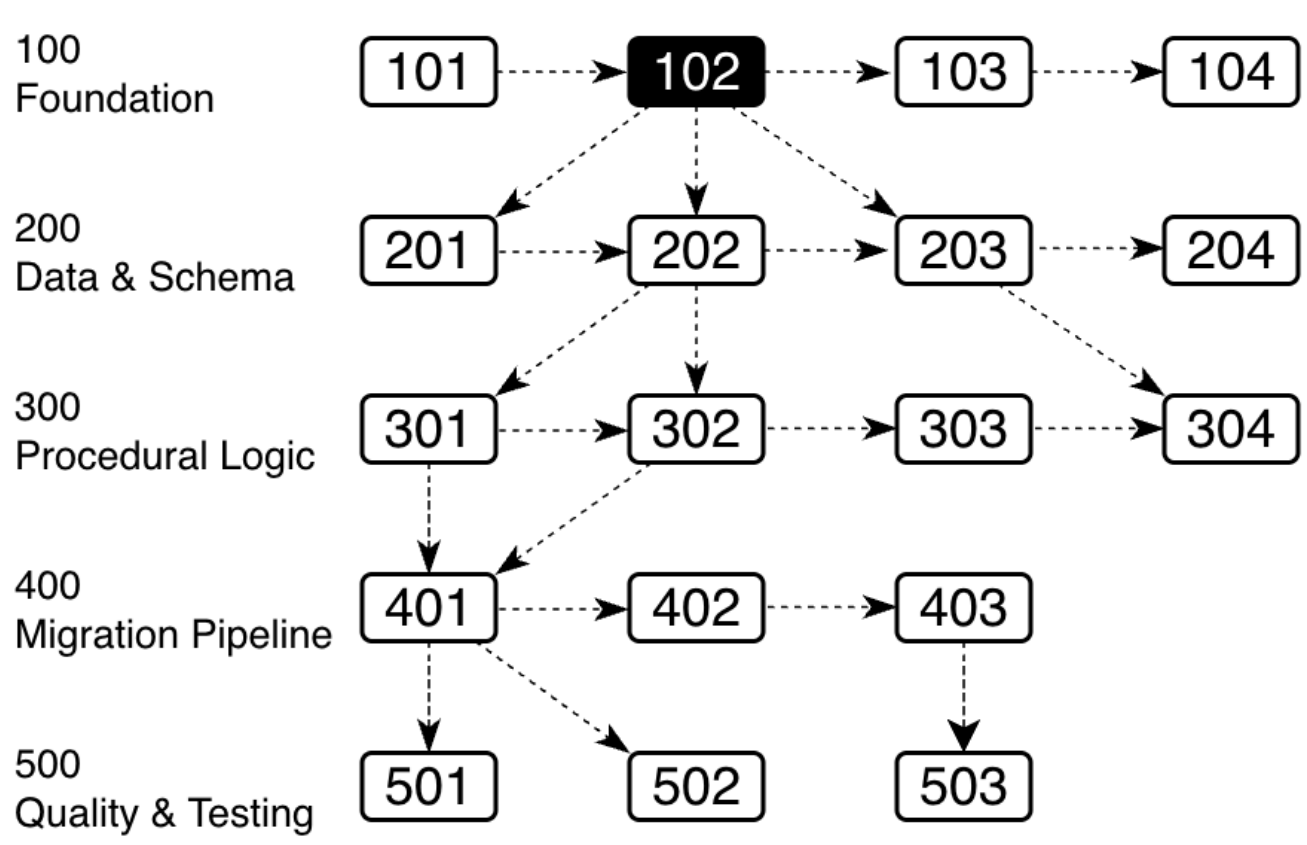

Fig. 1. Semantic-block dependency DAG; Block 102 propagates system constraints to all downstream blocks.

Table I summarizes the 18 semantic blocks and their principal decision ownership.

TABLE I. SEMANTIC BLOCKS AND REPRESENTATIVE DECISION OWNERSHIP

| Block | Title | What it owns (representative decisions) |
|---|---|---|
| **100 — Foundation** | | *project-wide constraints every lower block inherits* |
| 101 | Vision & Goals | business drivers, measurable success KPIs, project scope, cross-Oracle/PG glossary |
| 102 | System Constraints | target version (PG 16), extension **whitelist**, **forbidden** Oracle constructs (`DBMS_*`, DB links, hints), downtime budget, encoding — *the dominating block (W3)* |
| 103 | Security & RBAC | Oracle→PG privilege-model mapping, role/profile translation, RLS for VPD, `pgaudit`, encryption/TDE strategy |
| 104 | Architecture Decisions | ADR log: migration strategy (big-bang/phased), maintenance-window vs zero-downtime, tooling, DAL-change scope, dual-write |
| **200 — Data & schema** | | *static schema translation* |
| 201 | Datatype Mapping | Oracle→PG type matrix, `NUMBER(p,0)` banding (p≤9→`integer`, ≤18→`bigint`), `DATE`→`timestamp(0)`, `''`-vs-`NULL` policy |
| 202 | Tables & Indexes | DDL translation, constraint modes (`NOT VALID`), index redesign (bitmap→GIN), declarative partitioning, synonyms |
| 203 | Sequences & Identity | sequence vs `IDENTITY` strategy, the post-load `setval()` reset protocol, `NEXTVAL`-in-default rewrites |
| 204 | Collation & Charset | target encoding (UTF8), lowercase-unquoted identifier convention, `NLS_SORT`→ICU collation, `LIKE`/case-fold behavior |
| **300 — Procedural logic** | | *PL/SQL → PL/pgSQL behavior* |
| 301 | Package Decomposition | package→schema mapping, package-global-state workarounds (temp tables / GUCs), initialization blocks |
| 302 | Cursors & Loops | cursor-type translation, `BULK COLLECT`/`FORALL`, `ROWNUM`/`CONNECT BY` rewrites |
| 303 | Exception Handling | Oracle exception→SQLSTATE map, `RAISE_APPLICATION_ERROR`, mandatory error-logging pattern, `WHEN OTHERS` policy |
| 304 | Transaction Control | `AUTONOMOUS_TRANSACTION`→`dblink`, savepoints, `COMMIT`/`ROLLBACK`-in-procedure rules |
| **400 — Migration pipeline** | | *moving the data and cutting over* |
| 401 | Initial Load | tool selection (ora2pg), FK-aware load order, exclusions, pre-load validation |
| 402 | CDC Replication | change-capture tool, Oracle supplemental logging, lag thresholds, conflict resolution |
| 403 | Cutover & Fallback | step-by-step cutover runbook, sequence resets, measurable rollback criteria, `oracle_fdw` fallback |
| **500 — Quality & testing** | | *acceptance gates* |
| 501 | Data Integrity | row-count/checksum comparison methodology, acceptance thresholds, `DATE`/`NULL` re-validation |
| 502 | Functional Testing | test levels, per-procedure test template, non-negotiable go/no-go suite |
| 503 | Performance Benchmark | baseline capture (AWR), benchmark query set, PG tuning parameters, ≤15% regression threshold |

**Document anatomy.** Every document carries the same machine-parsable skeleton: a Status header, Purpose and Scope, the block's rules—concrete and actionable, with exact commands and decision tables rather than prose guidance—a Known Pitfalls section of named anti-patterns, an Open Questions register (Q-* identifiers with owner and due date), and a versioned revision history. Rules carry inline provenance: when a hardening loop changes a rule, the edit cites the gap that forced it, so the document history is an audit trail of $\Delta$ shrinking.

**Organizing principles.** Three principles connect the blocks. Reference, don't duplicate: every rule has exactly one owning document; other documents cite it; this makes W2 hold by construction and makes the coverage census of Section V-B well-defined. Constraint propagation: the foundation's system-constraints document (102) dominates—no lower document may contradict it. Ambiguity-stop: a document that does not determine a choice must record it as an explicit Open Question rather than leave silence—consumers are instructed to halt and flag, not guess.

A finding that belongs here rather than in the results. The latter two principles are what the specification claims makes 18 files behave as one. Section IX-E reports a controlled experiment in which they are supplied to implementers and nothing measurable changes. They remain load-bearing for authoring and review—they are what makes W2/W3/W4 checkable—but we can no longer claim they govern implementer behavior, and this paper does not.

**The dependency graph.** The structure is harvested mechanically from each document's Depends on: field, yielding a directed graph of 18 nodes and 19 edges—sparse (density 0.124), longest path 7, maximum level width 6, fan-in ≤ 2, fan-out ≤ 4—with unique root 101 and with 102 reaching every other block (Fig. 1; the same graph is navigable as an Obsidian vault, one node per block). Sparseness is a design outcome, not an accident: each new dependency edge grows some task's mandatory reading, so edges were admitted only where a document genuinely consumes another's decisions.

The practical consumption unit is the ancestor closure of a target block. For a task associated with block b, the required context is:

$$C(b) = \{b\} \cup Ancestors(b). \tag{3}$$

This allows the specification to be supplied in task-specific slices without manually selecting documents. Across the instantiated specification, the mean closure contains 4.72 documents, corresponding to 29.1% of the complete specification; hence dependency-based selection reduces mean task context by approximately 71% while retaining all declared dependencies.

This reduction should be interpreted as structural context selection rather than semantic compression: the retained documents are not summarized or rewritten. The dependency graph determines which complete specification blocks are required for a task.

The graph also provides the basis for formal verification and benchmarking. The same dependency declarations are used to generate the machine-readable model, verify structural properties, construct task-specific contexts, and record exactly which specification content each implementer receives. This makes specification structure part of the experimental object rather than merely a document-organization convention.

## IV. FORMAL MODEL AND WELL-FORMEDNESS

A specification is represented as:

$$S = (B, \prec, \rho, \Delta, Q), \tag{4}$$

where $B$ is a finite set of semantic blocks; $\prec \subseteq B \times B$ is the dependency relation; $\rho: B \to P(R)$ assigns rules to blocks; $\Delta$ is the set of decision points; and $Q \subseteq \Delta$ is the set of explicitly open questions.

A valid specification satisfies four well-formedness conditions.

**W1 — Acyclicity**:

$$(B, \prec) \; is \; a \; DAG. \tag{5}$$

A topological ordering must therefore exist.

**W2 — Single ownership**:

$$b \neq b' \Rightarrow \rho(b) \cap \rho(b') = \emptyset. \quad (6)$$

No rule is independently owned by more than one semantic block.

**W3 — Constraint domination.** There exists a foundation block F such that every dependent block remains consistent with its rules:

$$F \prec^{+} b \Rightarrow \rho(b) \textit{ is consistent with } \rho(F). \quad (7)$$

For the instantiated migration specification, F=102.

**W4 — Totality / ambiguity-stop.** Every relevant decision is either determined or explicitly represented as open:

$$\Delta = Det(S) \uplus Q. \quad (8)$$

Silence is therefore treated as a specification defect rather than an implicit authorization for an implementer to guess.

For the evaluated instance, W1–W4 are machine-checked. A graph-processing tool verifies W1 using Kahn's algorithm over the 18 blocks and 19 edges. W2 follows from the single-owner document hierarchy. W3 is checked against block 102, including target-version pins and extension restrictions. W4 is checked by an exhaustive scan of the documents: the specification contains 81 Q-* identifiers, no dangling references, an Open Questions register in all 18 documents, and no unflagged TBD/TODO markers.

An implementation M assigns a resolution M(δ) to each relevant decision δ∈Δ. It conforms to the specification,

$$M \vDash S, \quad (9)$$

when its decisions satisfy the rules assigned by ρ, including propagated foundation constraints.

A decision $\delta$ is determined by $S$ when every pair of conforming implementations resolves it identically:

$$S \vdash \delta \Leftrightarrow \forall M, M' \vDash S: M(\delta) = M'(\delta). \quad (10)$$

The model-theoretic determinacy of the specification is therefore:

$$det(S) = \frac{|Det(S)|}{|\Delta|}. \quad (11)$$

This definition separates determined decisions from explicit uncertainty. W4 requires that any decision not belonging to $Det(S)$ appear in $Q$.

The empirical difficulty is that observed implementer agreement does not necessarily imply that the specification caused the agreement. Strong implementers may independently select the same conventional solution even without a specification. Consequently, cross-implementer convergence is only an estimator of determinacy and requires a no-specification control to distinguish specification-induced agreement from shared implementation priors. This distinction motivates the benchmark design introduced in the following sections.

**Limitation of the empirical estimator.** The quantity $det(S)$ is defined over conforming implementations, whereas empirical convergence may also reflect shared implementer priors. Estimating it by implementer convergence silently assumes that observed agreement is caused by S. Section IX-B shows that this assumption fails in practice: on our corpus, every decision the specification pins is one on which implementers agree without it. The definition is unharmed—the estimator needs a denominator, which is the instrument's job.

# V. VALIDATING THE STRUCTURE CHOICE

## A. Closure economics

Before asking whether the specification is good, we ask whether its decomposition is defensible — a question usually settled by taste. Four computational checks, all reproducible from model.json and the corpora.

The full specification is 61,709 tokens. The mean per-task ancestor closure is 4.72 documents = 29.1% of the specification (worst case 42.9%): the structure reduces mean per-task context by approximately 71% while retaining all declared dependency-ancestor blocks. This is the structure's most practical payoff, and it is what makes the benchmark's task-matched context possible.

## B. Coverage census

The structural coverage analysis uses 4,254 anonymized production-derived units. Only construct-level statistics are retained; proprietary identifiers and business content are excluded. The corpus is used for structure validation and is distinct from the execution benchmark corpus. A study-defined taxonomy of 192 Oracle construct categories was derived from Oracle documentation. Across 4,254 anonymized production-derived units, the analysis identified 180,995 construct occurrences. Each observed occurrence was assigned to one semantic block, while the taxonomy-level census assigned 85.5% of the in-scope construct categories to a single block and identified 24 uncovered categories, which were triaged into seven families.

## C. Alternative partitions

Under a declared objective: "acyclicity + coupling + closure cost + width", the five-layer phase decomposition is a local optimum: no W1-satisfying alternative Pareto-dominates it, random same-granularity partitions are acyclic with $p \approx 0.0008$ and never beat it on coupling, and the ticket-literal concern grouping is fragile (78% of documents have multi-team ownership; a natural tie-break induces block-level cycles).

## D. Non-authored recovery

The layers are recovered at the 99.9th percentile from an edge set derived from document-body citations that the structure did not author—closing the circularity objection to the grouping claim. A fully exogenous edge set (construct co-occurrence within production units) has essentially no community structure (modularity ≈ 0.002), because a single production unit cuts across datatypes, cursors, exceptions, and transactions at once. That null does not refute the layering; it confirms the premise of A—migration tasks are inherently cross-layer, so what matters is closure cost, not code locality.

# VI. BENCHMARK FOR SPECIFICATIONS

## A. What is scored

Public SQL/LLM benchmarks score a model on fixed tasks. Here the implementer panel is fixed and the specification is the treatment; a run is a triple (specification version, unit corpus, panel). Every arm has a mandatory no-specification control: a determinacy number without its control is uninterpretable, because high agreement may be a shared industry default rather than a specification effect.

### B. *Task-matched context*

The context handed to an implementer is the computed dependency closure of the block levels the corpus exercises, derived from model.json per run and recorded in the run's metadata—never a hardcoded or inherited list. This is not a convenience: an early campaign inherited a closure from a previous study and withheld the entire 300 level from an 82%-procedural corpus. The resulting "97% determinacy on error codes" was an artifact—the specification had changed 0.0% of those decisions because implementers never saw the block that pins them. Provenance of what was sent is therefore part of every run record.

### C. *Deterministic judges, no LLM in the loop*

Two judges. Executability: each output is loaded into a pristine postgres:16 container and the first error is classified (ok / malformed / missing object / duplicate / other). Behavioral differential: the migrated unit is replayed on PostgreSQL and compared against the same unit's behavior captured on a live Oracle instance—error-class sequences (mapped through the specification's own exception table), and, in the newest instrument, structured `result_set` and `table_state` rows. Every verdict is reproducible from committed artifacts.

### D. *Corpus*

An internal 1,802-pair Oracle→PostgreSQL migration test corpus was execution-filtered before evaluation. Execution filtering revealed that only 981 of 1,802 candidate migration pairs executed successfully on both sides, showing that corpus validation is necessary before benchmark use. From the 981 we draw a nested, stratified sample of 200 whose 75-unit prefix is pinned as data. Behavioral work uses a separate corpus of 20 curated public PL/SQL units with permissive licensing, whose ground-truth behavior is captured on live Oracle.

### E. *Panel rules*

Two rules survive every panel decision. The panel must be identical across arms; otherwise, “what the specification changed” is confounded with “who ran it”. Any panel change is a deliberate, documented revision, never a mid-study swap. Seats are won on a fixed candidate probe—never on price, speed, or quota—because a weak implementer diverges from incompetence and the instrument would score that as specification ambiguity.

### F. *Nine non-redundant evaluation axes*

Executability; behavioral executability; determinacy; rule compliance (is a stated rule followed); coverage denominator (does the specification speak to this decision at all); behavioral fidelity; structured data; rule efficacy (does obeying a rule improve the outcome); and ambiguity honesty. We deliberately do not compute a weighted single score: a specification that breaks migrations would hide behind a high executability number, and Section IX-D exhibits exactly such a rule.

## VII. Pilot: Three Curated Schemas

The pilot study evaluated three curated schemas using three heterogeneous LLM implementers (Claude Opus 4.8, GPT-5.5, and DeepSeek-V4-Flash) and ora2pg, a deterministic migration tool that reads the Oracle catalog directly and does not receive the specification, on Oracle's official HR, OE, and SH sample schemas.

### A. *Convergence*

On every decision the specification determines, all three implementers and ora2pg agreed, across all three schemas—type-band mappings, DATE→timestamp(0), identifier folding, IOT→heap+PK, sequence semantics, declarative range partitioning (26 partitions), materialized views with query rewrite dropped, outer-join and function-based-index rewrites. A naive port control (Oracle DDL with only SQLPlus directives stripped) fails with 78 errors and 0 tables created, confirming that the successful loads cannot be explained by PostgreSQL accepting the unmodified Oracle DDL.

### B. *The hardening loop*

Twelve gaps were surfaced and attributed, including three hard defects that the engine rejected: NOT VALID emitted on a primary key; a foreign key with a numeric child against an integer parent; and ADD FK … NOT VALID on a partitioned table. Each was invisible to review, produced by exactly one implementer, independently compared with ora2pg's behavior, and eliminated by a one-rule change. This is the loop the model predicts: implementer disagreement is not noise but a measurement of $\Delta \setminus Det(S)$.

### C. *A powered ablation on arrangement*

90 runs ({HR, OE, SH} × {blocked, flat, knockout} × {GPT-5.5, DeepSeek-V4-Flash} × 5 repeats) tested whether arranging a task's closure into labeled, ordered blocks affects an implementer, holding content constant. Result: a null—the capable implementer is invariant across conditions (≤0.9 pp, unchanged even with the constraints block deleted), while the weaker implementer's run-to-run variance dominates any condition effect.

### D. *How to read this section in light of what follows*

Three curated schemas exercise a small, mostly static decision surface, and the 300 level is touched only incidentally. The pilot establishes that the loop works and that the defects it finds are real. It does not establish that convergence is caused by the specification — for that, a control arm and a coverage denominator are needed, and neither existed at pilot time.

## VIII. At Scale: What The Specification Determines

### A. *Three arms, identical units*

75 pinned units from the execution-verified corpus, two independent implementers, three arms differing only in the specification supplied: **none** (control), **partial closure** (the 200-level-only inheritance bug, retained as a free ablation), and **full closure**.

TABLE II. Executability and Aggregate Determinacy Across Specification Arms

| Arm | Executability (k=2) | Executability (k=3) | Aggregate determinacy (k=3) |
|---|---|---|---|
| no specification | 72.0% | 60.0% | 83.1% |
| partial closure | 87.5% | — | — |
| full closure | **97.3%** | **88.0%** | 83.8% |

**Executability is the specification's clearest, monotone win**—+25 pp on one panel, +28 pp on the other.

**Aggregate determinacy is effectively unchanged**— and on the second panel it is mildly negative. At k = 3 it moves

83.1 → 83.8%; at k = 2, on the same units, it moves **83.2 → 79.7%**. We report both because reporting only the favourable panel would be the exact error this paper warns about elsewhere. Implementers without any specification already agree ~83% of the time on shared industry defaults, so there is almost no agreement left to buy, and what movement exists is within the noise established in Section X. This result is specific to the evaluated contemporary implementer panel, not a specification failure — but it does mean the estimator proposed in Section IV does not discriminate at this scale.

### B. Structure beneath the null

Three patterns are visible even though the aggregate is flat. Partial closure *disperses* implementers (83.2 → 72.6 at k=2): an incomplete specification is worse than none, providing evidence that closure completeness affects cross-implementer consistency. Raw per-decision deltas include returns (+18.7 pp), routine_kind (+17.4 pp), and param_modes (+11.3 pp); however, the repeated-run analysis in Section X retains only the routine_kind effect as exceeding its axis-specific variability. Where it offers options rather than deciding, it *splits* implementers (`raise_levels` −17.8 pp). One axis, `type_set`, plateaus near 50%; decomposition shows this is a **metric ceiling**, not a verdict—roughly 20 of 75 units differ only on invented local variable types that no specification can determine, and the specification- accountable quantity (interface-level agreement) is 64–67%.

### C. "It loads" badly overstates "it works"

On the 20-unit PL/SQL corpus, executability is ~97% while **exact behavioral match against live Oracle is 2–6 units of 20**, with event-level fidelity 0.15–0.26. A load-or-agreement metric alone would score this class of code a near-total success. This gap motivates the behavioral evaluation lane. Among the SQL/LLM benchmarks reviewed in Section II [13], [14], [33], [34], we did not identify an equivalent Oracle-to-PostgreSQL behavioral differential based on live source-engine execution.

### D. Correctness of the data is unchanged

The newest instrument compares **structured rows** — `result_set` and `table_state` — of a migrated unit against a live Oracle. Alignment is at the level of engine-neutral **probes** drawn from each unit's own script, not statements: a migrated unit is not statement-isomorphic to its source (implementers merge, split, and reorder), so a positional diff would flag every reformatted unit. Normalization is declared, not silent, and reported alongside a strict byte-equal lower bound.

TABLE III. STRUCTURED-DATA COMPARISON AGAINST LIVE ORACLE

| | control | full specification |
|---|---|---|
| **probes matching live Oracle** | **19 of 42** | **19 of 42** |
| divergent (wrong rows/values) | 11 | 7 |
| genuine missing object | 2 | 6 |

The specification yields **exactly as many correct results as no specification**. What it changes is the *kind* of failure: four wrong-data divergences become outright missing objects. This was the first measurement that could have shown the specification helping at the data level, and it did not.

## IX. DESIGNED EXPERIMENTS

Six studies. Three are pre-registered manipulations with expectations fixed before any data existed (A, E, F); three are measurement axes built to interrogate the null of Section VIII rather than to test a hypothesis (B, C, D). Each isolates one variable.

### A. Rule placement affects compliance more than restatement in the tested condition

The rule text is byte-identical across arms; only its position in the datatype document differs. 31 pinned units, 3 repeats, k = 2, compliance measured as the share of string parameters rendered per the rule.

TABLE IV. RULE COMPLIANCE BY PLACEMENT CONDITION

| Placement | Compliance |
|---|---|
| absent (control) | 19% |
| trailing paragraph | 23% |
| **both places** | **34%** |
| **matrix row** | **42%** |

Two results. First, the matrix-row placement increases compliance from 23% to 42% (+19 pp)—a reversal of the naive reading we held before isolating the variable. Second, and unregistered by anyone: stating the rule in both places scores below the better single placement. The pre-registration allowed for "≈ max, not the sum" — the only anticipated surprise was compounding. In this experiment, duplicating the rule reduced compliance relative to its strongest single placement. A second, weaker statement does not reinforce the dominant signal; it competes with it. For specification authors, this result cautions against assuming that duplicate placement necessarily reinforces a rule.

The effect is carried by one implementer (38→78→47→64%); the other floors at 3–6% regardless of placement. The observed placement effect is therefore implementer-dependent and should not be generalized beyond the tested panel.

### B. A coverage denominator, and the null it exposes

For each decision on the surface we label whether the specification pins it, offers options, or is silent, then cross that label with agreement in both the full and no-specification arms. The full-vs-control delta separates "the specification caused this agreement" from "they would have agreed anyway."

On the evaluated 75-unit decision surface, no pinned decision demonstrates a measurable agreement gain over the no-specification baseline. Every one is either a shared industry default the specification would get for free (error codes: 100% agreement with or without the specification) or an outright compliance failure (parameter modes 47%, parameter types 53%). The single optioned rule disperses implementers (89% → 52%). Determinacy alone cannot see any of this: a rule everyone ignores in unison scores exactly like a rule everyone follows.

### C. Compliance as determinacy's required companion

Measured across specification versions, adherence to one pinned rule is non-monotone: control 13/26%, v0.1 33/2%, v0.2 57/20%, v0.2.1 50/3%. Both implementers comply best under the version we had logged as weaker, and worse after the rule was folded into the matrix. The v0.2.1 revision had been recorded as a success because splits fell — but they fell together with compliance. The models agreed more by ignoring the rule in unison. Every specification-version claim needs the compliance number beside the agreement number; this one corrected a verdict already in our own record.

### *D. A specification can make a migration worse*

Crossing compliance with behavioral outcome decomposes one collection rule into three claims that do not share a verdict. The rule's prose ("use a native array") is a would-agree-anyway default — implementers write arrays without it (control 15/24 vs 14/24 with). The rule's matrix example is what acts, and it acts destructively: it flips the element type (control 10 text / 0 varchar → 2/10) and the declaration carrier, moving both implementers from a valid CREATE DOMAIN … AS TEXT[] they write unprompted to an invalid CREATE TYPE … AS VARCHAR(100)[]. The control produces 0 invalid declarations; in this manipulation, the specification-induced rule choice produces 10 invalid declarations. On the compliance axis alone this reads as a specification success. This is the concrete reason Section VI-F refuses a weighted single score.

### *E. The specification's own consumption protocol is inert*

An audit established what implementers actually receive: the dependency graph is present (each block states its own Depends on, and all 19 edges are recoverable from the payload), but the specification's consumption protocol — constraint propagation and ambiguity-stop, the two principles of Section III — had never been supplied. Supplying it, with everything else frozen:

TABLE V. EFFECT OF THE SPECIFICATION CONSUMPTION PROTOCOL

| Measure | without protocol | with protocol |
|---|---|---|
| compliance | 42% | 39% |
| W4 precision (implementer A / B) | 0.352 / 0.374 | 0.350 / 0.374 |
| executability | 89.1% | 91.4% |

Compliance moves −2.3 pp against the observed median arm-delta spread of 14.4 pp measured in Section X. Ambiguity-log precision — registered as the outcome specifically because recall alone tracks log length — is flat to three decimals. Executability is unchanged, as pre-registered.

The consequence is a bound on what this programme may claim. "These documents determine decision X" holds. "The SDD *method* determines X" does not — and now because it was measured and did not appear, not because it was untested. The contrast with Section IX-A is notable: relocating one sentence changed compliance from 19% to 42%, whereas adding approximately 1,900 tokens of consumption guidance changed it by −2.3 pp.

### *F. In the tested repair task, execution feedback outperformed specification patching at approximately 1/14 of the inference cost.*

A pre-registered head-to-head: one repair round in which each implementer sees its own psql stderr ($0.03) cut broken-code errors 85 → 46 and confirmed both registered expectations; the comparable specification patch ($0.41) failed all three of its pre-registered expectations. The sharpened lesson, measured in both directions: specification patching was effective when the added rule arbitrated between plausible defaults, but ineffective when it contradicted a strong implementer prior. Specification and execution loop own disjoint divergence buckets — and a third bucket, roughly 60 events of missing behavior, is owned by neither. Its natural owner is a test that asserts the error fires.

## X. MEASUREMENT VALIDITY: EMPIRICAL RUN-TO-RUN VARIABILITY

Initial benchmark runs used one generation per experimental cell. LLM generation is stochastic; the same model, prompt, and unit produce different code on re-run. The programme had carried an unmeasured assumption of ±5 pp. We measured it: three repeats on a pinned 25-unit subsample, with the unit set held fixed across repeats (comparing one repeat on 25 units against another on the 12 that happened to finish measures the unit sets, not the churn — an error that inflated an early pass to 35 pp before it was caught).

TABLE VI. RUN-TO-RUN VARIABILITY OF THE MEASUREMENT INSTRUMENT

| | **max spread** | **median spread** |
|---|---|---|
| within the specification arm | 16.7 pp | 8.3 pp |
| within the control arm | 20.0 pp | 8.0 pp |
| **arm delta (what claims are about)** | **32.7 pp** | **14.4 pp** |

The delta spread is what matters, because claims are differences and the difference of two noisy quantities is noisier than either. One decision axis changes sign between identical re-runs (+23.3, −9.3).

**What this retires.** The floor is per axis, not global — which is why a smaller effect can survive while a larger one does not. Applying each axis's own measured delta spread.

TABLE VII. EFFECTS RELATIVE TO AXIS-SPECIFIC MEASUREMENT SPREAD

| Claim | Effect | That axis's spread | Verdict |
|---|---|---|---|
| rule pin (`param_modes`) | +3–5 pp | 32.7 pp | **retired** |
| `raise_levels` recovery | +6.7 pp | 16.5 pp | **retired** |
| `param_modes` (three-arm) | +11.3 pp | 32.7 pp | **retired** |
| `returns` (three-arm) | +18.7 pp | 20.3 pp | **retired** |
| `routine_kind` (three-arm) | +17.4 pp | 12.3 pp | survives |
| executability | +25 pp | ≈13 pp | survives |

If unit-level outcomes are approximately independent, the standard error of an agreement proportion decreases on the order of $1/\sqrt{n}$. Using this scaling only as a heuristic, the 25-unit repeated subset suggests variability on the order of 8 pp at n=75. The returns effect (+18.7 pp) is retired, whereas routine_kind (+17.4 pp) exceeds its measured axis-specific spread. Reporting a single global threshold would have gotten both of these backwards. An independent corroboration arrived from Section IX-E, where executability varied 12.9 pp across identical re-runs of the same arm.

Under the present three-repeat design, effects below approximately 8 pp at n=75, or 14 pp at n=25, should not be interpreted as robust specification effects without additional repetitions. We state this as a limitation of the instrument, not of the specification, and we report it because a benchmark that cannot bound its own noise cannot adjudicate anyone else's claims — including ours.

## XI. DISCUSSION

### A. What a specification demonstrably buys

Three things survive measurement. Executability: +25 pp of clean-loading output, monotone across arms and panels, exceeding the observed executability run-to-run spread. Checkability: W1–W4 are mechanically verified, and the closure structure cuts per-task context by 71%. Gap localization: divergence can provide a localization signal when it persists beyond observed run-to-run variability — the pilot's twelve gaps, the census's 24 triaged holes, and the honesty axis all converge on the same open questions.

### B. What it does not buy, on the evaluated contemporary LLM implementers

Agreement. Implementers already concur on shared industry defaults, and every decision our specification pins turns out to be such a default or a compliance failure. This is the paper's central negative result, and it is a statement about the era as much as about the artifact: an instrument built to detect specification-induced agreement finds little to detect for the evaluated implementer panel when the decisions align with common migration defaults.

### C. Where the leverage actually is

In the tested placement manipulation, rule location produced a larger compliance effect than restatement. The same rule, byte-identical, increases compliance from 23% to 42% from the document's dominant structure as from a trailing paragraph — and stating it in both places is worse than stating it once. Meanwhile the added meta-level consumption guidance produced no measurable effect in the tested protocol manipulation. For practitioners the guidance is concrete: put a pin in the structure a reader is already using, once, and do not hedge it with a restatement.

### D. Determinacy after the null

We do not withdraw the definition. det(S) remains a coherent model-theoretic notion of what a specification fixes, while W4 provides an explicit treatment of unresolved decisions. What the measurements retire is the estimator's discriminating power in this regime, and the naive reading that high convergence evidences a good specification. The denominator is the fix: agreement is evidence about a specification only relative to what implementers would have done without it.

### E. On reporting negative results

Four of this paper's contributions are things that did not work, and one is the retraction of our own earlier claims. We report them because the alternative — an instrument whose authors publish only its confirmations — cannot be trusted to measure anyone else's specification either. These reversals show that the measurement protocol can falsify previously favorable interpretations rather than only confirm them.

### F. Property-specific and evolutionary scope

The present results should not be generalized into a single scalar notion of specification quality. Security-oriented SDD shows that explicit knowledge can affect a different downstream property [6]–[8], while state- and drift-oriented studies treat software and monitoring requirements as context-dependent and evolving [9]–[11], [35], [36]. Our benchmark evaluates one frozen specification state; whether its determined decisions remain adequate after architecture, policy, context, or goal changes is not measured here.

## XII. THREATS TO VALIDITY

**Shared blind spots (Knight–Leveson).** LLM implementers are not statistically independent; unanimous agreement can overstate determinacy. Mitigated by a non-neural baseline (ora2pg), an execution judge, and the coverage denominator of Section IX-B, which converts the threat into a measurement and is precisely what revealed the null. Not eliminated: independence was verified only as pairwise disagreement (the panel's least-agreeing pair at 78%).

**Single case, fictional estate.** One specification, one migration direction, one demo estate of 18 documents. The evaluation uses an internal execution-filtered migration corpus and public Oracle sample schemas, but the specification governs a scenario written for the study. Whether the placement effect and the protocol null transfer to a production specification of hundreds of documents is untested, and is the most important open question for the programme.

**Model era.** Every result is conditioned on the specific model versions and provider endpoints evaluated in this study. The determinacy null in particular is a statement about implementers that already share strong defaults; on weaker or more heterogeneous implementers the same specification could show a large effect. The pilot's weaker implementer already hints at this.

**Statistical power.** Three repeats provide an initial estimate of run-to-run variability but do not characterize its full distribution. Effects in approximately the 8–15 pp range should be treated as exploratory unless supported by additional repetitions.

**Instrument limitations, stated rather than hidden.** The structured-data comparison carries two measured caveats, identical across arms so that neither moves the headline: a fixture asymmetry between the two engines affecting four probes per arm, and six probes that are Oracle-only and therefore classified unjudged rather than failed. The behavioral corpus is 20 units. The executability judge classifies the first error only.

**Specification and agent evolution.** The fixed-panel design is an experimental control, not a production orchestration prescription. Operational multi-agent systems may recruit specialized agents dynamically [37], and evolving systems may require specification or monitoring updates as architecture and context change [10], [11], [37]. Such changes would alter both the treatment and implementer population and require a separate longitudinal causal design.

**Cost as a confound on scope.** The whole programme cost ≈ $74 of inference. Several arms were sized by budget rather than by power — notably the second protocol arm, which was pre-registered as conditional and not run once the first returned a null.

## XIII. CONCLUSION AND FUTURE WORK

This study formalizes specification structure and evaluates specification determinacy through controlled independent implementation. The formal answer holds: a specification is $(B, \lessdot, \rho, \Delta, Q)$ with four machine-checkable conditions, its five-layer decomposition is computationally defensible rather than merely tasteful, and determinacy is a well-formed model-theoretic notion of quality.

The empirical answer is more interesting than the one we expected. For the evaluated contemporary LLM implementers, the full specification increased clean-loading output by 25 percentage points on the primary two-implementer comparison while implementer divergence and coverage analysis provide observable signals for gap localization—but it does not produce a measurable aggregate agreement improvement in the evaluated control comparison, while structured-data correctness remains unchanged at 19 of 42 matching probes. In the tested placement manipulation, rule location affected compliance, while duplicating the rule reduced compliance relative to its strongest single placement.

Future work follows from the threats rather than from the successes. Scale the evaluation to a production specification and a production migration estate—the fictional-estate threat is the binding one, and it is also where retrieval versus full-closure context becomes a genuine question rather than an unnecessary confound. Reps on the remaining axes—repeated-run variability was characterized on the decision surface only; executability, compliance, and the behavioral channels each need their own. Close the missing-behavior bucket with tests that assert the error fires, the one divergence class that neither specification nor execution feedback owns. And turn the placement result into author guidance, since it is the finding most likely to be useful to someone who will never run this benchmark.

A further direction is longitudinal specification benchmarking. A temporal extension $S_t = (B_t, \prec_t, \rho_t, \Delta_t, Q_t)$. Such an extension could test whether determinacy and rule efficacy persist under architectural, policy, contextual, security, or goal drift. This extension is motivated by evolving-system and specification-driven monitoring studies [10], [11], [35], [36] but is not evaluated in the present experiments.